\documentclass[11pt,twoside]{article}
\usepackage{asp2014}

\begin{document}

\title{Stellar Cosmic Rays in a Habitable Zone}
\author{Alexei Struminsky$^{1,2}$ and Andrei Sadovski$^{1,2}$}
\affil{$^1$Space Research Institute, Moscow, Russia; \email{astrum@iki.rssi.ru, asadovsk@iki.rssi.ru}}
\affil{$^2$Moscow Institute of Physics and Technology, Dolgoprudniy, Russia}

% This section is for ADS Processing.  There must be one line per author.
\paperauthor{Alexei Struminsky}{astrum@iki.rssi.ru}{ORCID_Or_Blank}{Space Research Institute}{Plasma Physics Department}{Moscow}{Moscow}{117997}{Russia}
\paperauthor{Andrei Sadovski}{asadovsk@iki.rssi.ru}{ORCID_Or_Blank}{Space Research Institute}{Plasma Physics Department}{Moscow}{Moscow}{117997}{Russia}

\begin{abstract}
According to recent observations relative number of flare stars does not change very much from cool dwarfs to hot A stars. Flare energies are strongly correlated with stellar luminosity and radius.  Whence it follows that the typical magnetic field associated with a flare is several tens gauss and the typical flare loop length-scales are parts of the stellar radius. Flares on O--B stars were not observed, but they are possible, since strong magnetic fields are detected on O--B stars. Therefore stars of O--M spectral classes are potential sources of cosmic rays. Energy estimates of a magnetic field strength in a tube in photospheres of O--M stars are performed. Basing on their values possible flare energies and numbers of accelerated protons are estimated. The values obtained for the Sun correspond to observations by order of magnitude that justify estimates for other stars.  Values of magnetic field strength in a tube differ less than five times for O and M flares (700 and 3500~G), but corresponding flare energies and numbers of accelerated protons for O stars are greater by five orders. Contrary fluencies of stellar protons appear to be five orders less.
\end{abstract}

\section{Introduction}
Stellar flares are considered as a possible source of galactic cosmic rays (GCR) nearly since the discovery of solar cosmic rays. In spite of the long history the question is not resolved yet (\cite{Unsoeld57,Lovell74,Mullan79,KopysovStozhkov05}). The near sources of GCR might be responsible for the anomalous PAMELA effect (\cite{Stozhkov11}).  Stellar cosmic rays (SCR) are the important factor of space weather in a habitable zone near cool dwarfs (\cite{Tabataba16}). Stellar sources of CR across Hertzsprung--Russell (H--R) diagram have not been considered.  

If these stellar sources have the same nature as solar flares then they should be associated with enhanced magnetic and coronal activity. The most massive coronal main sequence(MS) stars are late-A or early-F stars, a conjecture that is supported both by observation and by theory (\cite{Gudel04}). In earlier--type stars of spectral type O and B shocks developing in unstable winds are the likely source of X-rays. Theory predicts the absence of a magnetic dynamo, given the lack of a significant outer convection zone in A-type stars. The outer convection zones of stars become very shallow toward early F stars and disappear in A-type dwarfs. Even if magnetic fields existed in early A-type stars, efficient coronal energy release is not expected because no strong surface convective motions are present to transport energy into non-potential coronal fields (\cite{Gudel04}). 

The statistical analysis of flares across H--R diagram observed by the Kepler space telescope (\cite{Balona15}) showed that the relative number of flare stars decrease from about 10\% for K--M stars to 2.5\% for A--F stars. Surprisingly, there does not seem to be a drop in stellar activity as the granulation boundary is crossed.  In a case of M and K stars the rate of energy release during the flare is of the same order as the normal release of the star in the quiescent condition, therefore they have showed, possibly, a better statistics of observations.  Flare energies are strongly correlated with stellar luminosity and radius, that the typical magnetic field associated with a flare is several tens of gauss and the typical flare loop length-scales are parts of the stellar radius (\cite{Balona15}). The flare activity of O--B type stars might not be excluded, but the flare power should be small in comparison with stellar luminosity.

There are a number of excellent reviews of stellar magnetic field across H-R diagram (\cite{Berdyugina09,DonatiLandstreet09,Linsky15}). Average surface magnetic field in most cases roughly equal to the field strength at which the magnetic field balances thethermal pressure of surrounding gas. The equipartition magnetic field strength is the minimum possible value for the field strength dynamic flare loops. Massive stars with strong magnetic fields and fast rotation are very rare and pose a mystery for theories of star formation and magnetic field evolution (\cite{Hubrig15}). The magnetic properties of O and early B type stars are extencively investigated  by different collaborations, for instance (\cite{Wade16}).

 The goals of this paper are to estimate ultimate numbers of cosmic rays accelerated in stellar flares across the H--R diagram and evaluate their fluencies in a habitable zone.

\section{Ultimate flare energy and stellar cosmic rays}

Equating average magnetic field stellar surfaces to the field strength at which the magnetic field balances the thermal pressure of surrounding gas, $B^2/8\pi=nkT=\rho GMH/{R^2}$, and in assumption that characteristic scale is equal to the mean free path $H=\lambda=(n\sigma_T)^{-1}$ ($\sigma_T$ is the Thompson cross-section),
we obtain the estimation of the photospheric magnetic field strength
$B_{ph}=B_{0\odot}(R_\odot/R)(M/M_\odot)^{1/2}$,
where index $\odot$ is used for the definitions of corresponding Sun values.

Results of our calculations for MS stars across H-R are presented in Table. According to observations the typical magnetic field associated with a flare 
is several tens of gauss and the typical flare loop length-scales are parts of the stellar radius (\cite{Balona15}). Therefore, the flare energy may be 
estimated as ${B^{2}}/({8\pi})L^{3}=\beta^{2}\alpha^{3}{B_{ph}^{2}}/({8\pi })R_{\ast }^{3}$, where $B=\beta B_{ph}$, $\beta<1$, $L=\alpha R$, $\alpha<1$, $R_{\ast }$ is the 
stellar radius.
For the flare energy we get an estimate $E_{fl}=2.3\times{10}^{37} \beta^{2}\alpha^{3}(R/R_{\odot})(M/M_{\odot})$~erg.

Let us estimate the ultimate number of charge particles $N$ with average energy $E$, which might be accelerated in the active region with such values of $B$ and $L$. We get
$N=1.4\times10^{42}(\beta^{2}\alpha^{3}RM)/(ER_{\odot}M_{\odot})$~protons with average energy $E$ (MeV).
Here we suppose that particles are stored in the magnetic trap during the 
acceleration process and released simultaneously. 
Below relative values of flare energy and accelerated proton number for 
different spectral classes would be interesting for us, therefore results 
presented in Table are normalized to $\alpha^{3}\beta^{2}$ ($\alpha 
=1$, $\beta =1$).

For simplicity we suppose that the habitable zone is at a such distance $H$ from 
the star, where a flux of stellar luminosity is equal to the solar one at 
the Earth orbit (1~AU). A fluence of stellar CR at a distance of habitable 
zone $H$ in a case spherically symmetric propagation is $f=N/(4\pi EH^{2}(1.5\times{10}^{13}~\text{cm})^{2})$.

\begin{table}[t]
	\caption{}%The different stars parameters}
	\tabcolsep=2pt
	\begin{center}\small
		\begin{tabular}{lcccccccccccc}
			\hline
			Class & 
			${L}/{L_{\odot}}$& 
			${M}/{M_{\odot}}$& 
			${R}/{R_{\odot}}$& 
			${g_{\ast}}/{g_{\odot}}$& 
			$E_{fl}/E_{fl\odot}$& 
			$\tau/\tau_\odot$& 
			$H$, AU& 
			$B_{ph}$ G& 
			%Number of protons $\alpha^{3}\beta^{2}$, 
			$N$, 30~MeV& 
			%Proton Fluence $\alpha^{3}\beta^{2}$ cm$^{-2}$
			$f$, 30~MeV \\
			\hline
			O5& 
			$7.9\times10^5$& 
			60& 
			14& 
			0.31& 
			$826$& 
			0.001& 
			889& 
			710& 
			$3.9\times10^{43}$& 
			$1.7\times10^{10}$ 
			\\
			\hline
			B0& 
			$5.2\times10^4$& 
			16& 
			7.4& 
			0.29& 
			$117$& 
			0.002& 
			228& 
			693& 
			$5.6\times10^{42}$& 
			$3.8\times10^{10}$\\
			\hline
			B5& 
			830& 
			7& 
			3.9& 
			0.45& 
			$27$& 
			0.033& 
			29& 
			870& 
			$1.3\times10^{42}$& 
			$5.6\times10^{11}$ \\
			\hline
			A0& 
			54& 
			3& 
			2.4& 
			0.52& 
			$7.4$& 
			0.14& 
			7.3& 
			925& 
			$3.4\times10^{41}$& 
			$2.3\times10^{12}$\\
			\hline
			A5& 
			14& 
			2& 
			1.7& 
			0.69& 
			$3.4$& 
			0.24& 
			3.7& 
			970& 
			$1.6\times10^{41}$& 
			$4.1\times10^{12}$ \\
			\hline
			F0& 
			6.5& 
			1.8& 
			1.5& 
			0.80& 
			$2.7$& 
			0.41& 
			2.5& 
			1147& 
			$1.3\times10^{41}$& 
			$7.3\times10^{12}$ \\
			\hline
			F5& 
			3.2& 
			1.5& 
			1.4& 
			0.76& 
			$2.1$& 
			0.65& 
			1.8& 
			1122& 
			$9.9\times10^{40}$& 
			$1.0\times10^{13}$ \\
			\hline
			G0& 
			1.5& 
			1.05& 
			1.1& 
			0.87& 
			$1.1$& 
			0.75& 
			1.2& 
			1195& 
			$4.7\times10^{40}$& 
			$1.2\times10^{13}$ \\
			\hline
			G5& 
			0.8& 
			0.92& 
			0.92& 
			1.0& 
			$0.83$& 
			1.03& 
			0.89& 
			1338& 
			$4.0\times10^{40}$& 
			$1.8\times10^{13}$ \\
			\hline
			K0& 
			0.4& 
			0.78& 
			0.85& 
			1.1& 
			$0.65$& 
			1.63& 
			0.63& 
			1468& 
			$3.1\times10^{40}$& 
			$2.7\times10^{13}$ \\
			\hline
			K5& 
			0.15& 
			0.69& 
			0.72& 
			0.96& 
			$0.48$& 
			3.19& 
			0.39& 
			1480& 
			$2.3\times10^{40}$& 
			$5.3\times10^{13}$ \\
			\hline
			M0& 
			0.08& 
			0.51& 
			0.6& 
			1.42& 
			$0.30$& 
			3.80& 
			0.28& 
			1527& 
			$1.4\times10^{40}$& 
			$6.3\times10^{13}$ \\
			\hline
			M5& 
			0.01& 
			0.2& 
			0.27& 
			2.74& 
			$0.05$& 
			5.22& 
			0.1& 
			2125& 
			$2.2\times10^{39}$& 
			$7.7\times10^{13}$ \\
			\hline
			M8& 
			0.001& 
			0.1& 
			0.11& 
			8.26& 
			$0.01$& 
			10.9& 
			0.03& 
			3352& 
			$5.1\times10^{38}$& 
			$2.0\times10^{14}$\\
			\hline
			\end{tabular}
			\label{tab1}
			\end{center}
\end{table}

\section{Discussion} 

The obtained values of photosphere magnetic field roughly correspond to observations (see fig.~1 in \cite{Berdyugina09}). If such magnetic field exist in the photosphere then a creation of stable magnetic field structures (arcs or loops) with typical scale of parts of stellar radius are possible. These loops may interact near their tops resulting in flares and other coronal activity. Only a small fraction of massive stars ($\sim7${\%} for O--B stars) seem to host a measurable structured magnetic field, whose origin is still unknown. It is widely assumed that hot O--B stars do not have well developed convection zone and, consequently, the well-established flare activity. 

The statistical analysis of Kepler flares across H--R diagram perfomed by \cite{Balona15} showed that the relative number of flare stars decrease from about 10\% for K--M stars to 2.5\% for A--F stars. A ratio of the estimated flare energy to the stellar luminosity is a characteristic time ($\tau \alpha^{3}\beta^{2}$ in Table, for the Sun $\tau_\odot\alpha^{3}\beta^{2}\approx17754$~s), which is necessary for considerable input of flare energy to the total stellar luminosity. This time is very short for hot massive stars and, possibly, explains why flares are not observed at O--B stars. These flares might be responsible for coronal heating and strong stellar wind. Observations of X-ray emission with high time resolution may help to detect flares at O--B stars. 

The calculated values of possible flare energy (consequently, proton accelerated number) differs by about five orders for O and M stars. Accounting values of $\alpha $ and $\beta $ may smooth this difference. The magnetic field strength determined from the median flare energy is a few tens of gauss, which is quite reasonable in the framework of the magnetic reconnection model (\cite{Balona15}). Therefore $\beta =0.1$ might be assumed for all stars. A depth of the convection zone determines dimension of active regions. Therefore $\alpha $ parameter should be larger for cool dwarfs, where a depth of the convection zone may reach a half of stellar radius and a spot group may cover a half of visible hemisphere. If the flare region covers a half of visible hemisphere, then $\alpha =1.77$. The ultimate flare energy for cool dwarfs may increase by $\alpha^{3}=5.6$ times. This may lead to equality of ultimate flare energies for cool and hot stars. Contrary a difference between fluencies of SCR will increase.

Let us check how realistic are our estimates of ultimate flare energy and proton fluences for the Sun. The famous 775~AD event of radionuclide increase recorded in tree rings, the largest solar proton event recorded so far (\cite{Kovaltsov14}), requires $>30$~MeV proton fluence $F$30 of $8 \times 10^{10}$~$\text{proton}\cdot\text{cm}^{-2}$, a solar source of such event should be the flare with total energy $9 \times 10^{33}$~erg covering 16000~mhs ($16000\times10^{-6}\times2\pi =\alpha^2$, $\alpha =0.37$), the soft X-ray class is $\sim X230$ (\cite{Cliver14}). According to our estimates we get smaller values of $F=\alpha^{3}\beta^{2}5\times10^{14}\times1~\text{MeV/E}$~$\text{proton}\cdot\text{cm}^{-2}/E$, if protons propagate in 4$\pi $, but quite reasonable $\alpha^{3}\beta^{2}1.8\times10^{16}\times1~\text{MeV/E}$~$\text{proton}\cdot\text{cm}^{-2}/E$ within 36 times smaller angle, i.\:e. $2\times10^9$--$3\times10^{11}>30$~$\text{MeV proton}\cdot\text{cm}^{-2}$ ($\alpha$ is varies from 0.1 to 0.37, $\beta =0.1$). The values obtained for the Sun correspond to observations by order of magnitude that justify estimates for other stars. 

\section{Conclusions}

Energy estimates of a magnetic field strength in a tube in photospheres of O-M stars are performed. Basing on their values possible flare energies and numbers of accelerated protons are estimated.

The values obtained for the Sun correspond to observations by order of magnitude that justify estimates for other stars. 
Values of magnetic field strength in a tube differ less than five times for O and M stellar flares (700 and 3500~G), but corresponding flare energies and numbers of accelerated protons for O stars are greater by about five orders. 

Fluencies of stellar protons appear to be five orders less for O stars in comparison with M stars in their habitable zone.

\acknowledgements The work was partly supported by the Russian Foundation for Basic Research (grant 16-02-00328) and the Programm 1.7 P2 of the Russian Academy of Sciences.

\end{document}